# Magneto-dielectric coupling and non-ergodic electrical behaviour in hexagonal $Sr_{0.6}Ba_{0.4}MnO_3$ via local strain driven magnetic ordering

Ritu Rawat[a], R.J. Choudhary[a]*, A.M. Awasthi[a]*, Rajamani Raghunathan[a], Archna Sagdeo[b], A.K. Sinha[b], S. Chaudhary[c], S. Patnaik[c], and D.M. Phase[a]

[a]UGC DAE Consortium for Scientific Research, Indore- 452 001, India

[b]Indus Synchrotrons Utilization Division, Raja Ramanna Centre for Advanced Technology, Indore- 452 013, India

[c]School of Physical Sciences, Jawaharlal Nehru University, New Delhi 110067, India

**Abstract**

The crystal structure of hexagonal-$Sr_{0.6}Ba_{0.4}MnO_3$ allows various competing superexchange interactions, leading to intriguing magnetic properties. Local structural changes modify overlapping between Mn and oxygen ions with temperature. Calculations based on our model spin-Hamiltonian reveal that the dominant linear antiferromagnetic superexchange interaction between the oxygen-linked $Mn^{4+}$ ions results in short range correlations (SRC), manifesting a smooth drop in magnetization below 325K. Dominance of superexchange interaction changes its allegiance towards the non-linear oxygen-linked Mn-O-Mn interactions, onsetting long-range correlations (LRC) below 225K. Below the SRC-LRC crossover temperature, electrical response arising from the interacting dipoles exhibits power-law divergent behaviour of relaxation time, upon cooling. Non-ergodic character of the dipole-cluster glass state is examined via the indispensable aging and rejuvenation effects, similar to the spin glasses. Competitive-frustration among spin-exchange and local-strain is reckoned as responsible for the electrical glass origin.

*Corresponding Authors: ram@csr.res.in, amawasthi@csr.res.in

In recent years' magneto-electric (ME) materials have attracted tremendous attention owing to their spectacular applications in various fields such as spintronics, sensors, 4-state memory device, etc [1]-[4]. These materials display the coupling between the magnetic and electrical properties. Understanding the microscopic mechanism of ME-coupling is crucial from fundamental as well as applications perspective. In some materials, generally ME-coupling is ascribed to the coexisting magnetic and electrical ordering, though with independent origin of orderings. On the contrary, in some materials, ME-coupling is triggered upon the contingent origin of magnetic and electric orderings. In recent years, a new type of ME coupling is reckoned in systems with either or both ordering as short-ranged [5]. This has opened new avenues for different materials, which were otherwise not considered as multiferroic, given their spatio-temporal inversion-symmetry characters. Included amongst such are $YbFe_2O_4$, which shows simultaneous freezing of spin and polar degrees of freedom, indicating a strong ME coupling [6], multiglass $SmFeO_3$[7], and antiferromagnetic quantum paraelectric glass $SrCu_3Ti_4O_{12}$ [8].

Cubic $SrMnO_3$ follows Pm-3m symmetry with linear 180° Mn-O-Mn superexchange interaction leading to G-type AFM interaction with Nèel temperature ~230K [9]. It is shown that strain applied on cubic $SrMnO_3$ by chemically doping with Ba induces ferroelectricity due to displacement of Mn ion at ~400K [9]. In the paramagnetic regime at high temperatures, strain deviates the Mn-O-Mn angle from 180°, which restores to 180° on PM to AFM transition, hence high ME-coupling is observed [10]. Hexagonal (*h*-) $SrMnO_3$ follows $P6_3/mmc$ symmetry with linear 180° Mn-O-Mn and non-linear (< 90°) Mn-O-Mn superexchange interaction. *h*-$SrMnO_3$ also orders G-type antiferromagnetically, with $T_N$ ~280K [11]. However, strained hexagonal-$SrMnO_3$ is not well explored for its magnetic and electrical properties. In this study, we explore the coupling among structural, magnetic, and electrical properties of the hexagonal-$Sr_{0.6}Ba_{0.4}MnO_3$ (SBMO), which follows $P6_3/mmc$ symmetry, similar to its parent compound. Magnetic properties appear to be governed by intricate exchange interactions among the $Mn^{4+}$-ions within $Mn_2O_9$ bi-octahedra (via face-shared oxygen) and in the adjacent bi-octahedra (via corner-shared oxygen). It is observed that local structure parameters viz., Mn-O bond length, Mn position etc. vary non-monotonically with temperature, giving rise to magnetic anomalies, concurrent with the non-ergodic electrical behavior, displaying ME coupling in the system.

Polycrystalline hexagonal-$Sr_{0.6}Ba_{0.4}MnO_3$ (SBMO) was prepared using solid state reaction method. Magnetization measurement was carried out using 7T SQUID VSM (Quantum Design, USA). Dielectric measurements over 10Hz to 1MHz were performed in the temperature range of 100K to 430K in cooling cycle using Alpha-A high performance frequency analyzer (Novo Control). Temperature dependent XRD measurements were performed using Synchrotron source at RRCAT Indus-2, BL-12. Temperature dependent Magneto-dielectric measurements are performed using Agilent E4980A LCR meter in O and 5T magnetic field. Field dependent isothermal magneto-dielectric measurements at 1kHz$|_{210K}$ were performed using Alpha-A high-performance frequency analyzer (Novo Control) and a 9Tesla Integra cryostat/magnet (Oxford NanoSystems).

XRD analysis of the SBMO sample discussed in the Ref. [12] reveals its single phase nature with hexagonal $P6_3/mmc$ symmetry.

Dielectric constant $\varepsilon'(T)$ (Fig. 1(a)) shows a frequency dependent shoulder (marked as P1) like feature below ~ 230K. In $d\varepsilon'/dT$ plot (Fig. 1(b)), two sets of peaks are noticed; one with a systematic $\omega$-$T$ dispersion, corresponding to the shoulder (feature P1 in Fig. 1(a)) and another, relatively frequency indifferent one around ~320K. Dispersive peaks appear more clearly in loss tangent $\tan\delta_\omega(T)$ shown in Fig. 1(c), shifting towards higher temperature with increasing frequency. Above ~227K, magnitude and dispersion of $\tan\delta(T)$ change abruptly, as illustrated by the locus of the peak-maxima. Note that the typical "flat loss" spectral feature [13] is not observed here, which characterizes the relaxor ferroelectrics viz., the merging of loss tangent curves of different frequencies on their lower-temperature side. Therefore, the data apparently rule out the signature of polar nano regions (PNR's)/relaxor state.

For a precise determination of the electrical relaxation character, fits on the frequency-domain $\tan\delta_T(\omega)$ were performed (Fig. 1(d)), using general Havriliak-Negami (H-N) expression for the complex permittivity [14]:

$$\varepsilon^* = \varepsilon' - i\varepsilon'' = \frac{\Delta\varepsilon}{[1+(i\tau\omega)^\alpha]^\beta} - i\left(\frac{\sigma_0}{\varepsilon\omega}\right)^n + \varepsilon_\infty \quad (1)$$

Where $\varepsilon^*$ is the complex permittivity, $\varepsilon''$ is imaginary permittivity, $\Delta\varepsilon$ is known as dielectric strength of the material, $\tau$ is the mean relaxation time, $\sigma_0$ is d.c. conductivity, and ($0 \leq \alpha, \beta \leq 1$) parameterize the width (broadening) and shape (asymmetry) of relaxation peak spectra, respectively. Optimized $\alpha, \beta \neq 1$ from the H-N fits to tan$\delta_T(\omega)$ spectra mark their asymmetry (Fig. 1(d)) and *T*-dependent/extra-Lorentzian FWHM (inset), affirming the non-Debyean nature of the observed relaxations [14]. The non-Debyean character clearly discounts any significant extrinsic contributions to the dielectric response due, e.g., to the Maxwell-Wagner type inter-granular free charges, or to the hopping of intra-granular oxygen-ions. Such independent and uncorrelated polaronic degrees of freedom represent extraneous and conduction-dominant responses, which may feature generic Lorentzian relaxation at very low frequencies, corresponding to the basic Debyean exponential time-decay dynamics [15].

Accurate $\tau(T)$ obtained from the optimum H-N fits is plotted as ln($\tau$) versus *T* in Fig. 2. Clear change of curvature in ln($\tau$)-*T* at 220K reveals two types of relaxation mechanisms (I and II). Region II indicating low temperature divergence fits the characteristic power-law for cluster glasses, borrowed from critically-diverging dynamics of spin-glasses [16];

$$\tau = \tau_o \left(\frac{T-T_g}{T_g}\right)^{-zv} \qquad (2)$$

Here $\tau_o$ is the approach time for the nascent (just nucleated) dipole-clusters, $T_g$ is the glass transition temperature, and $zv$ is dynamic exponent. Best fit yielding $\tau_o$ = 10.79μs, $T_g$ = 100.83K, and $zv$ = 7.39 is shown in Fig. 2. Optimized dynamic critical exponent is comparable to that for BaTi$_{0.65}$Zr$_{0.35}$O$_3$ (BTZ35) [17], and for the 3D magnetic dipolar glass LiHo$_{0.045}$Y$_{0.955}$F$_4$ [18]. Rising sluggishness $\left(\tau \xrightarrow[T \to T_g]{} \infty\right)$ of dynamically cooperative glassy clusters owes to their diverging size-scale $\left(\xi \xrightarrow[T \to T_g]{} |T - T_g|^{-\nu}\right)$, since the critical relaxation time itself scales with the correlation length as $\tau \sim \xi^z$ [19, 20]. Taking $z \approx 3$ for reasonably isotropic (bulk) correlations— comparison with typical relaxor-PNR's (size ~$O$(nm) and approach time ~$O(10^1)$ ps) estimates the cluster size here as $\xi_{SBMO}$ ~$O(10^2)$ nm— the mesoscopic scale. The anomalous behavior of $\tau(T)$ in region I (Fig. 2), bending over to saturation on the lower-*T* side— also witnessed in (Gd, Eu, Tb, Dy)MnO$_3$ by Schrettle et. al. [21]— indicates local quantum tunneling mechanism [22].

We have directly witnessed the essential aging and rejuvenation attributes of the electrical cluster glass phase (region II in Fig. 2) by performing the isothermal waiting experiments below and above $T_g$. Following the standard protocol similar to as employed e.g., in the literature [23,24], our results for the dielectric constant measured at 15Hz over 50-160K are shown in Fig. 3. We chose the low probing frequency, so the bigger magnitudes of both the dielectric constant and the changes in it, measured over the temperature window of interest, provided a good signal to noise ratio for the aging effects as well. We first note that the uninterrupted baseline runs (① & ②) at uniform temperature-ramps (±0.5K/min, solid & dash line curves in the main panel) reveal a clear hysteresis between the cooling and warming data, along with their excellent overlaps at the lower and upper temperature-ends. This anomalous & novel feature is hitherto unreported, to the best of our knowledge, and its reproducivity has been duly verified by repeated cycles. Secondly, the isothermal aging/annealing in otherwise uniformly cooled run (③; open stars in the main panel) involved several hours' waiting and data collection each at three temperatures viz., ~ $1.1T_g$, $0.9T_g$, and $3/4T_g$ (lower inset). Clearly measurable ~$O(10)$% decays in dielectric constant were recorded in the isothermal experiments. Fractional falls during aging at these selected temperatures below the uniform cooling baseline are shown in the upper inset. Asymptotic rejuvenations back to the uniform cooling baseline (to within ~2%) in the post-isothermal/succeeding cooling laps are similar to those reported for the electrical cluster glasses [25] and for spin glasses [26]. Also shown in main panel is the uniformly warm up data (④; open circles in the main panel) taken immediately after the cooling run interrupted with the three isothermal waiting. Surprisingly, in this special warm up run, we did not encounter any detectable trace of the dielectric 'holes' created at the isothermal waiting temperatures while cooling. This anomalous absence of the so-called 'memory' effect has been reproducibly confirmed over repeated such cycles. The excellent overlap of post-aging (circles) and generic (dashes) warm-up runs' data, over the full temperature range of observation is remarkable. We did not find a repeat cooling run ⑤ measurably different from the original baseline uniform cooling ①, and the same is not shown here for clarity. Essentially, all the same features, albeit having rather low signal-to-noise ratios, were also observed at 110Hz probing frequency.

Consistent with the sub-$T_g$ non-ergodicity (inability of glassy state to visit all its dynamically sluggish metastable states on finite timescale), relatively largest (upper inset) & slowest (lower inset) isothermal-aging-drop in $\varepsilon'$ are seen during waiting at 91K (i.e., just below $T_g$). Consolidation of dynamical freezing with further lowering of temperature (with still possible thermal activation of yet-fewer/incompletely-frozen/less-sluggish metastable states) causes a lower drop at ~$3/4T_g$. We are given to understand the 'no memory' (warm up run, ④) signals witnessed here as consistent with fairly large and fast isothermal aging attribute— decays completed at the waiting temperatures within $\sim O$(hr) (lower inset, also see [32]), along with the rapid rejuvenations on reverting back to the cooling lapses of the run ③. As per discussed in the next sections, witnessed increase & decrease of the Mn-O bonds' inequivalence signifying high flexibility of the local structural adjustments and vibronic attributes— are the definitive convict for the observed features of our aging/rejuvenation/memory findings.

Temperature dependent XRD patterns (150K-330K) were recorded using synchrotron source. XRD patterns reveal that the crystal symmetry remains $P6_3/mmc$ down to the lowest measuring temperatures. However, Ba-doping leads to inequivalent elongation of Mn-O and Mn-O1 bond length, which give rise to the dipole moment in SBMO. In Fig. 4(a) and (b) we have plotted the difference in Mn-O and Mn-O1 bond lengths ($\Delta$) and shift in the Mn-position along the $z$-direction versus temperature, respectively. It is evident that $Mn_z$-position initially drops with temperature and below 240K (approximately near the P1 feature in dielectric), the shift of $Mn_z$-position changes its direction. This $\Delta z$ turnover of Mn-position reflects the rise-back of bond-length inequivalence (Fig. 4(a), leading to buildup of dynamic dipolar correlations. We denote the temperature range below 225K as the glass region (GR). Shifts in the $Mn_z$-position also lead to concurrent variations in the Mn-O-Mn bond-angle (Fig. 4(c)).

The variations in the microscopic structural parameters also modulate the magnetic properties of SBMO. The unit cell (*u.c.*) of this 4H structure has two face-sharing $Mn_2O_9$ bi-octahedra; which in turn link together in a corner-sharing fashion through a common oxygen (Fig. 5(a)). The corner-shared linkage leads to a 180° antiferromagnetic superexchange (AF-SE) ($J_1$) in the network. Further, within the bi-octahedra, the short $Mn^{4+}$-$Mn^{4+}$ distances (~2.50 Å) and the ~80° $Mn^{4+}$-O-$Mn^{4+}$ triads through the face-shared oxygen atoms lead to direct- ($J_D$) and super-exchange ($J_S$) interactions respectively, giving rise to an effective magnetic exchange $J_2$. The actual strength and sign of the magnetic interactions $J_1$ and $J_2$ will depend on the extent of the orbital overlap between the magnetic ion and the intervening oxygen [47].

From our magnetic data, as shown in Fig. 5(b) we observe that below 325K, the magnetization starts decreasing gradually; albeit the width of the transition is rather broad. Below 225K ($T_M$) however, magnetization drops even more gradually, with a discontinuity in its slope.

Below both these temperature values, *M-H* behavior (shown in Fig. 5(c)) is linear pointing towards the AFM nature of the sample in the studied temperature regime.

Though our temperature dependent XRD experiments show no significant change in the *u.c.* dimensions along magnetic anomalies observed, the $Mn_z$-position and Mn-O-Mn bond angles show clear anomalies across $T_M$. The Mn-O-Mn bonds within the bi-octahedra undergo a sharp variation close to $T_M$; the Mn-O-Mn bond lengths change by about 0.026 Å (~1.34%), and angles show an overall ~1.5° change.

The structural information presented above suggests that sharp changes in the magnetic exchange interactions are plausible close to 225K. The Heisenberg spin Hamiltonian, being an effective model, does not describe such competing superexchange and direct interactions adequately. However, a rudimentary way to model this is by fitting the magnetization data, using a spin-Hamiltonian separately in the two temperature regimes namely; (1) $T > 225K$ ($T_M$), wherein the antiferromagnetic interaction between the $Mn^{4+}$ ions of the linear Mn-O1-Mn bonds leads to short range correlations (SRC) and, together with extremely weak interactions between $Mn^{4+}$ ions of non-linear Mn-O-Mn bonds, give rise to a broad feature at 325K, and (2) $T < 225K$ ($T_M$), where the local structural changes sharply increase the intra-bioctahedral Mn-O-Mn antiferromagnetic interactions, resulting in long-range correlations (LRC).

The magnetic data can be modeled by considering the Hamiltonian,

$$\hat{H} = -J_1 \hat{S}_2 \cdot \hat{S}_3 - J_1 \hat{S}_4 \cdot \hat{S}_1 - J_2 \hat{S}_1 \cdot \hat{S}_2 - J_2 \hat{S}_3 \cdot \hat{S}_4 \quad (3)$$

With $J_2 \ll J_1$ for $T > 225K$ ($T_M$) and $J_2 \gg J_1$ for $T < 225K$ ($T_M$), where, $J_1$ ($J_2$) corresponds to the strength of inter (intra) -bioctahedral exchange interaction, $S_i$'s are the spin operators, and the subscripts on the spin operators correspond to the indices of $Mn^{4+}$ sites in a *u.c.* The interaction between sites '4' and '1' imposes the periodic boundary condition in the *u.c.* Positive (negative) values of $J_i$ correspond to ferromagnetic (antiferromagnetic) interactions. The model Hamiltonian in eq.(3) conserves both total $\hat{S}^2$ and $\hat{S}_z$ operators and hence it is possible to construct the Hamiltonian matrix (*H*) either in total-*S* or in total-$M_S$ basis. However, in this case the *H*-matrix is constructed in constant-$M_S$ basis, and then the eigenstates $E(S, M_S)$ of the spin model are numerically obtained by full diagonalization [48]. The expectation values of the $\hat{S}^2$ and $\hat{S}_z$ operators for each eigenstate are obtained, from which the total spin $S$ and $M_S$ of the state are deduced. Finally, the magnetization of the system as a function of temperature ($T$) at a chosen magnetic field ($H_z$) is obtained from the relation,

$$M(H,T) = N_A g \mu_B \frac{\sum_S \sum_{M_S} M_S Exp[-\{E(S,M_S) - g\mu_B H_z M_S\}/k_B T]}{\sum_S \sum_{M_S} Exp[-\{E(S,M_S) - g\mu_B H_z M_S\}/k_B T]} \quad (4)$$

where, $N_A$ is the Avogadro number, $\mu_B$ is the Bohr magneton value, $g$ is the Landè $g$-factor (taken to be 2.0), and magnetic field $H_z$ is set to 0.01 Tesla, as per the experiments. The magnetic data in the temperature range 225K-350K is fitted by iterating over the values of $J_1$ and $J_2$ (Figure 5(b)). Our best fit for the magnetization data in this temperature region yields $J_1$ = -18 meV and $J_2$ = -2.16 meV. The large values of $J_1$ suggest that the Mn ions are dimerized along the linear Mn-O1-Mn bonds, while the small magnitude of $J_2$ signify that in the intermediate temperature regime (225-350K), thermal energy overcomes this exchange coupling strength, leading to short-range ordering of Mn ions along linear Mn-O1-Mn bonds.

Now, we turn our focus on regime-2, corresponding to temperature values below 225K ($T_M$). Here, we consider the enhanced non-linear Mn-O-Mn superexchange via the face-shared oxygen plus the direct $Mn^{4+}$-$Mn^{4+}$ exchange ($J_2 \gg J_1$), that facilitates long-range ordering in the system. This is supported by temperature dependent neutron diffraction studies on similar systems [13], which showed long-range AFM ordering of both intra- and inter-bioctahedral $Mn^{4+}$ spins below 270K. The Hamiltonian (3) is once again solved by the method discussed above, with $J_1$ fixed at -18 meV. The magnetic data is fitted via iterative process, by allowing only $J_2$ to vary, with the condition that $J_2 \gg J_1$; our best fit yields $J_2$ = -86 meV. The strong antiferromagnetic nature of the effective interaction $J_2$ vis-à-vis $J_1$ can be understood in terms of the contributions from both (enhanced) super- and direct- exchange mechanisms between the $Mn^{4+}$ ions within each bi-octahedron. Unlike the previous temperature regime wherein the weak inter-dimer interactions forbid long-range ordering, here in this regime the strong inter-dimer AFM intereactions ($J_2 \gg J_1$), due to local structural changes, is the key for the onset of long-range ordering.

It should be noted that previous first-principles calculations based on the density functional theory had shown that there is no significant direct overlap of charge densities between the two $Mn^{4+}$ ions of the bi-octahedron [49]. However, it should be noted that these calculations were based on ground state structures, and hence the local distortions taking place versus temperature were not taken into consideration. In a related previous study on magnetic properties of 4H-SBMO, the SRC's at high temperatures were assigned to antiferromagnetic fluctuations within the bi-octahedron, and the low temperature properties were described based on the onset of LRC's due to the inter bi-octahedral interactions [50]. On the contrary, our experimental results on the crystal structure, magnetic measurements and supported by our theoretical model, show that the broad hump occurring at 325K is a result of antiferromagnetic fluctuations between $Mn^{4+}$ ions of two *adjacent* bi-octahedra, and the long range ordering sets with the sudden increase of magnetic exchange *within* the bi-octahedron, as a result of the local structural distortion.

To further confirm that below 225K a different magnetic arrangement takes over, we performed the magnetization versus field measurements from 200K to 300K at 5K interval. From these measurements, we plotted the slope of *M-H* curves versus temperature (Fig. 5(d)), which shows an anomaly at 225K. Thus, our experimental results and model calculations throw light on the strength of the magnetic exchange mechanisms involved in SBMO. Neutron diffraction experiments as well as a microscopic electronic model that includes electron-phonon interactions should deepen our understanding of the complex magnetism in this system, which needs to be further explored.

It should be recalled here that the broad transition at 325K coincides with the first peak in $d\varepsilon'/dT$, signaling a magneto-electric (ME) coupling. Furthermore, appearance of the frequency-dispersive peaks in tan$\delta$ coincides with the magnetic SRC to LRC transition at $T_M$, clearly evidencing segmental electrical-organization (size-distributed entities' response), triggered by the long range correlation of spins. Sub-$T_M$ shifting of Mn-ions results in increase of Δ, consequentially strengthening the dipole moment, which indicates the entanglement of spins & dipoles in SBMO. From theoretical calculations, it is shown above that the non-linear Mn-O-Mn superexchange interaction ($J_2$) is dominating below $T_M$. However, there continues a significant contribution from the linear Mn-O1-Mn super-exchange interaction ($J_1$). Therefore, a competition between the two exchange interactions leads to frustration in the dipole ordering, prompting the origin of the electrical cluster glass. The micro-structural parameters like $Mn_z$ position, Mn-O-Mn bond angle as well as Δ, reveal anomalous behavior at $T_M$, signaling a coupled structural, electrical, and magnetic behavior. Singh et. al. showed for $0.9BiFeO_3$-$0.1BaTiO_3$ composite, that while the structure remains the same, change in the bond-angle across the magnetic transition temperature and hence, shifts in the position of atoms, attribute to the ME coupling [43].

Magneto-dielectricity MD(%) is given by [51] $\frac{\varepsilon'(H)-\varepsilon'(0)}{\varepsilon'(0)} \times 100$

MD% versus temperature is shown in inset of Fig 6. It is observed that MD% is negative and its value is in the range of 0-1.2%. With decrease in temperature below 240K, MD% slightly increases, leading to a mild peak at 225K near the magnetic ordering temperature $T_M$, further confirming magneto-electric coupling in SBMO. We measured MD(*H*) under magnetic field up to 5T at 210K/1kHz, shown in Fig. 6; MD$_{210K}$(*H*) is all-negative valued. To discern if and how the (magneto) conductance influences our results, we have also plotted the magneto-loss [51] $\left\{ML(\%) = \frac{\tan\delta(H)-\tan\delta(0)}{\tan\delta(0)} \times 100\right\}$ at 210K in Fig. 6. Linear MD(*H*) and low-lying/flat ML(*H*) (solid lines) yielding |MD/ML| ~(6-11)dB unambiguously ascertain robust and genuine magneto-dielectricity up to 5 Tesla field.

Summarizing our work, it is established that the ME-coupling manifest in SBMO is due the structure-conserving/local-variation of Mn-O bond lengths in the MnO6 octahedra. Experimental and theoretical modeling reveals short-range correlations above 225K, where the linear Mn-O1-Mn superexchange interaction between the adjacent bi-octahedra's $Mn^{4+}$ ions dominates. Below 225K, enhanced magnetic exchange interaction between the intra-octahedral $Mn^{4+}$ ions is found to set in the long-range spin-ordering. Electric dipole moment, arising due to the inequivalent Mn-O bond lengths in the MnO6 octahedra, is entangled with the spins. Long-range spin correlations having contribution from both $J_1$ and $J_2$ lead to frustration in the dipole alignment, which gives rise to electrical cluster glass state. The latter is well confirmed by its characteristic aging and rejuvenation features. Unlike *c*-SBMO, *h*-SBMO does not show long range ferroelectric ordering, however, both reveal negative MD% below AFM long range ordering. While the *c*-SBMO reveals long range AFM ordering below 180K, *h*-SBMO reveals long range AFM ordering below 227K and be yond that short range AFM ordering exists, even up to 600K. Such variation arises due to the difference in their crystalline structure, which provides different Mn-O-Mn pathways for magnetic interactions. From our study, SBMO comes across as one of the rare multifunctional materials.

**Acknowledgments:** Authors are thankful to Dr. N.P. Lalla and Ms. Poonam Yadav of UGC-DAE CSR, Indore, for providing zero-field dielectric data.

**Figure Captions:**

**Fig. 1** (a) Dielectric constant $\varepsilon'(T)$ at selected frequencies. Inset shows field emission scanning electron microscopy of SBMO. (b) shows $d\varepsilon'/dT$ at the similar frequencies (c) $\tan\delta(T)$ at different frequencies and (d) $\tan\delta(\omega)$ at several temperatures (solid lines are H-N fits), inset shows the peaks' full width at half maximum FWHM$(T)$.

**Fig. 2** Ln($\tau$) versus $T$; solid line fit in region II illustrates glassy behaviour while the inset shows Ln($\tau$) versus Ln($T/T_g$-1), with solid straight line as the fit.

**Fig. 3** Aging measurements of dielectric constant at 15Hz frequency and ±0.5K/min temp-ramp rate under the following protocol— ① baseline uniform cooling (solid curve), ② baseline uniform warming (dashed curve), ③ uniform cooling intercepted by isothermal waiting/annealing at $T_w$ = 1.1$T_g$, 0.9$T_g$, and 3/4$T_g$ (open stars), and ④ uniform warming immediately post the cooling/aging run (open circles). Upper inset— fractional drops in permittivity at the annealed temperatures and its rejuvenation back towards the baseline (uniform-cooling) values, during the sectional-cooling lapses after the isothermal annealing. Lower inset— exponentially-decaying time profiles of $\varepsilon'_{T_w}$-drops at the annealing temperatures $T_w$, over the isothermal waiting times $t_w \sim O$(Hrs).

**Fig. 4** (a) Rietvield refinement of synchrotron XRD, performed across magnetic transition temperature at 210K and 300K, (b) difference between Mn-O1 and Mn-O bond length ($\Delta$), plotted versus temperature (solid line is guide for eyes), (c) shift in Wyckoff position of Mn along the $z$-direction with respect to temperature, and (d) shows change in bond angle Mn-O-Mn versus temperature.

**Fig. 5** (a) Schematic of the magnetic model, showing the pathways of magnetic exchange and the corresponding exchange strengths in LRC and SRC regimes. The oxygen atoms of linear and non-linear Mn-O-Mn bonds are marked in different colors for clarity. (b) Magnetization *vs*. temperature fits (continuous lines) of SBMO in SRC and LRC regimes. The experimental values are shown as red circles. The magnetization curve computed in either of these two regimes is extended into the other, to emphasize the change in slope across the magnetic anomaly. Inset shows high temperature magnetization up to 600K. (c) *M-H* behavior at 175K and 300K. (d) Slope of the *M-H* curves recorded at different temperatures across $T_M$; dashed line-- guide to eye.

**Fig. 6** Magneto-dielectricity MD(%) and magneto-losses ML(%) versus the applied field $H$ at 1kHz/210K. Inset shows MD(%) versus temperature at 1MHz/5T.

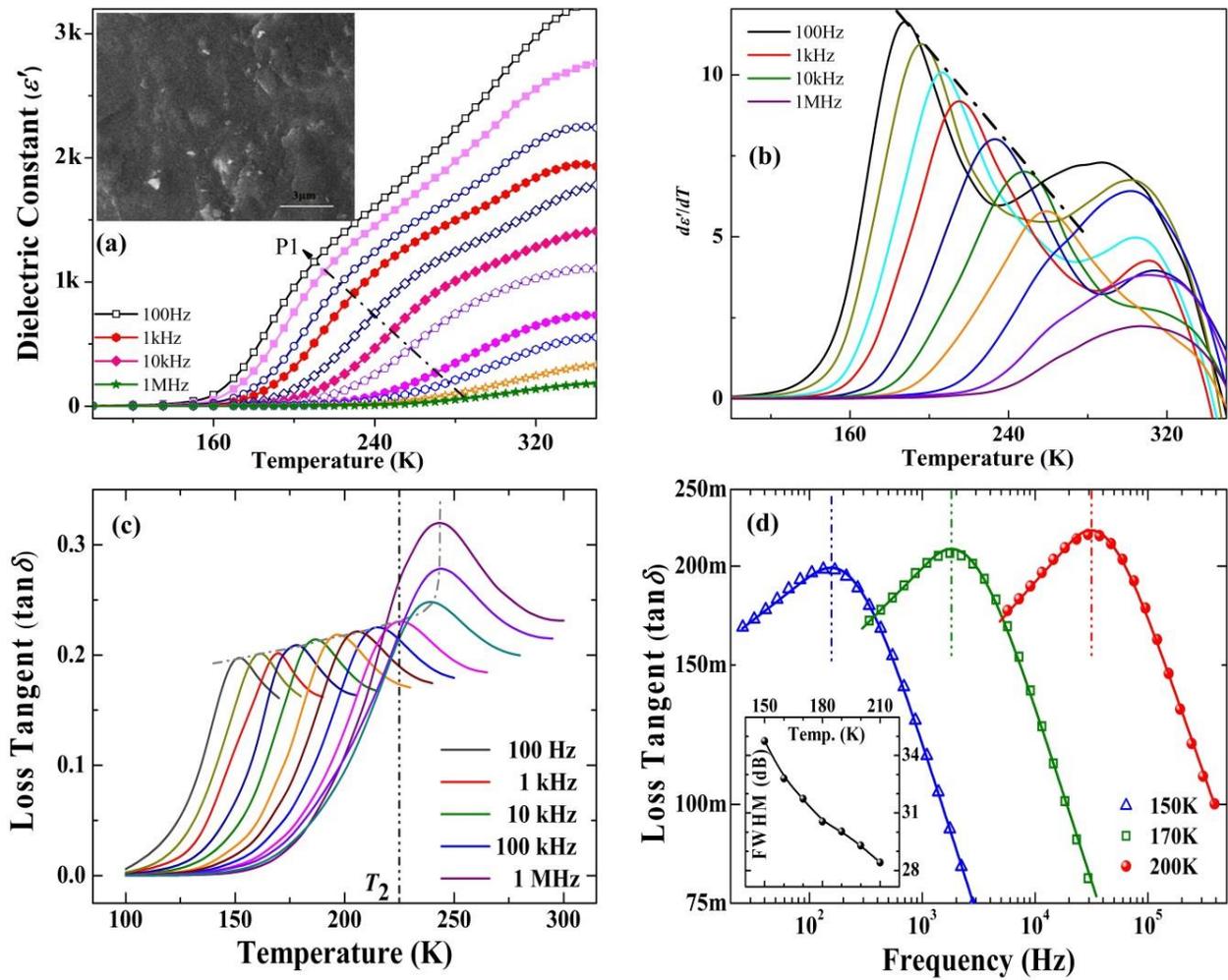

**Fig. 1**

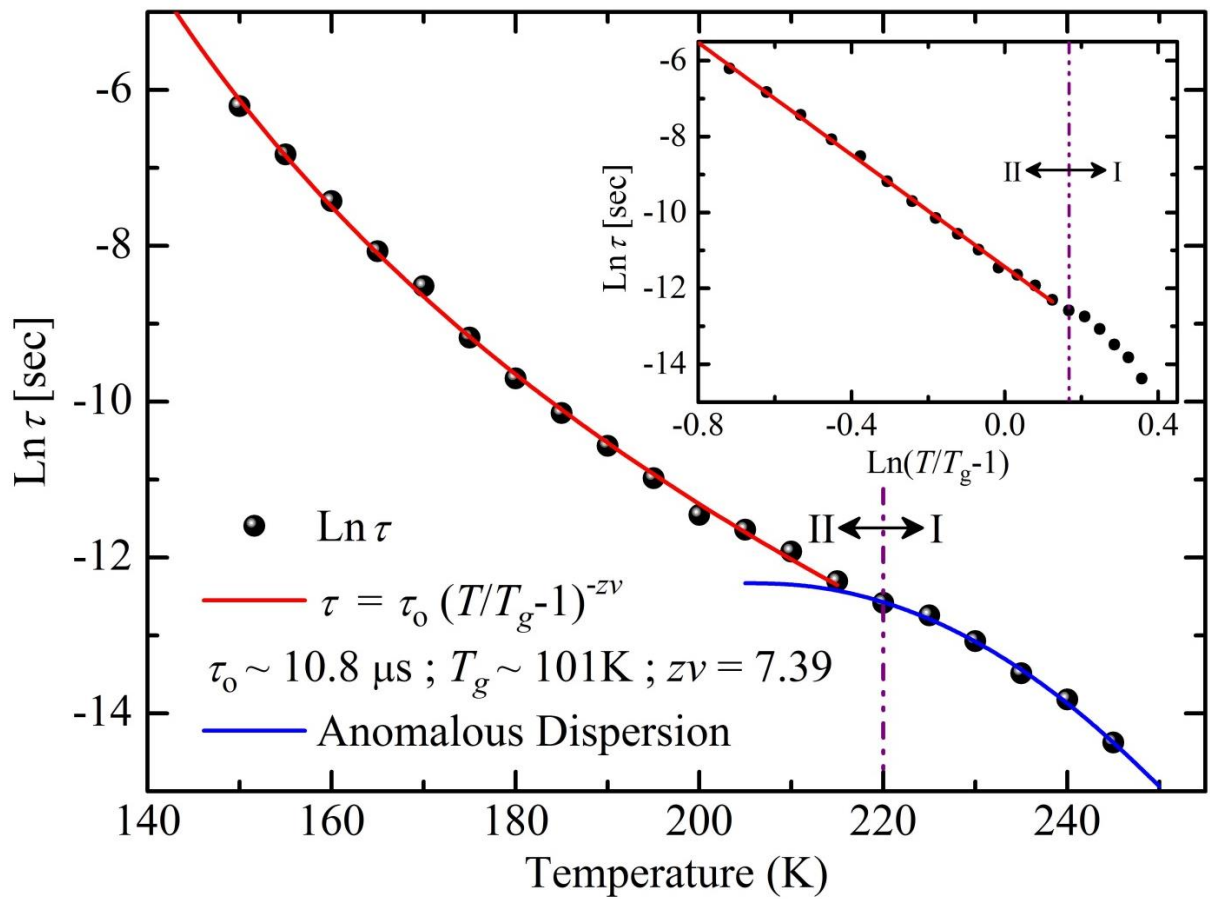

**Fig. 2**

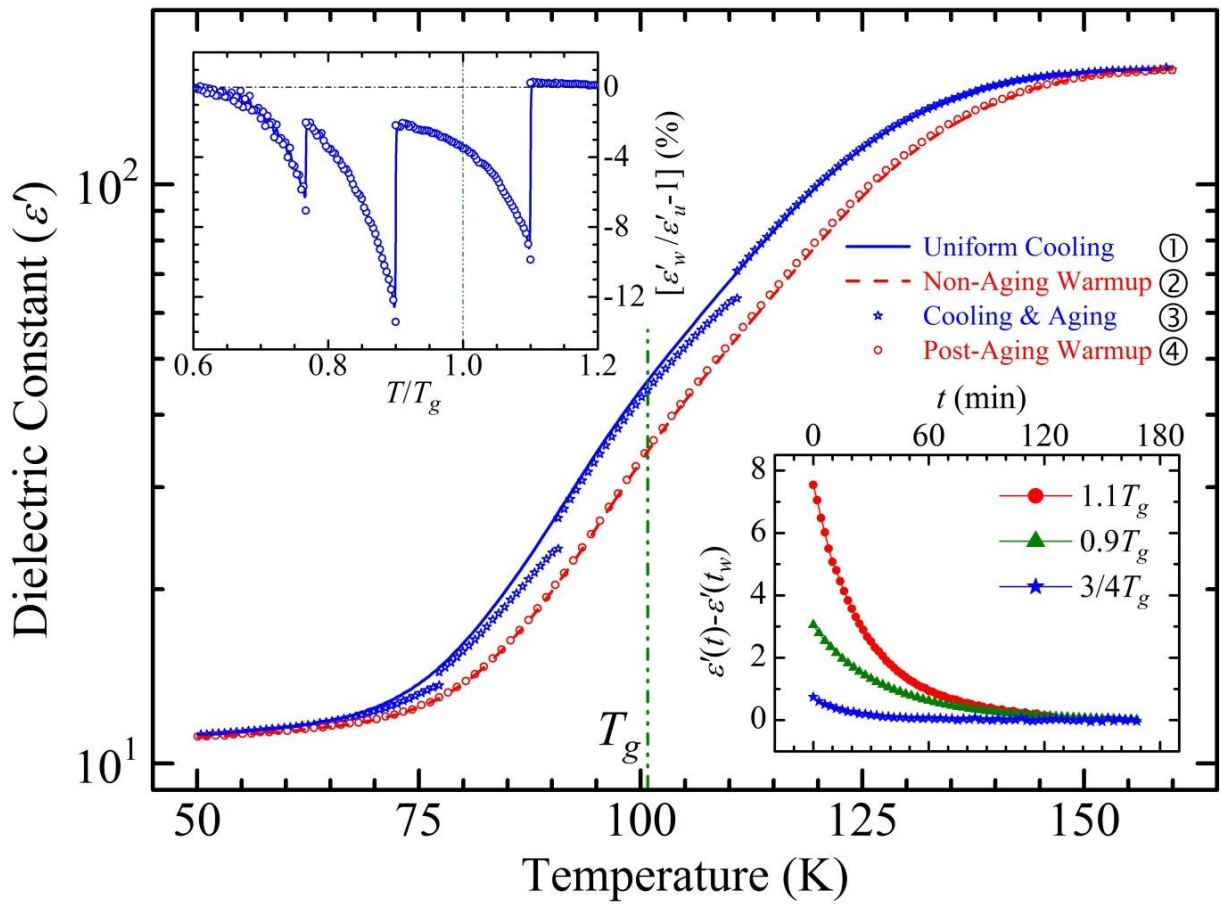

**Fig. 3**

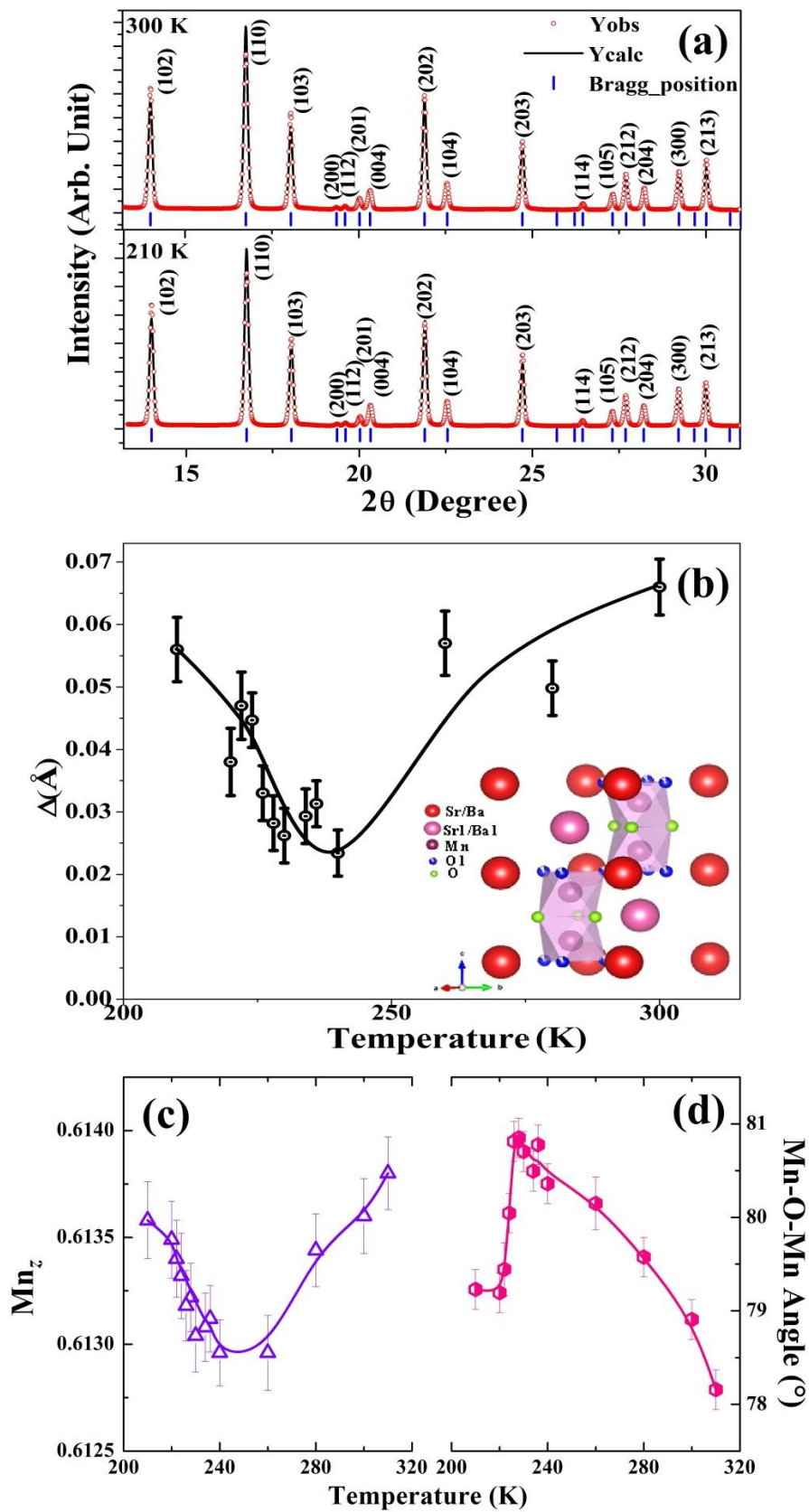

**Fig. 4**

Fig. 5

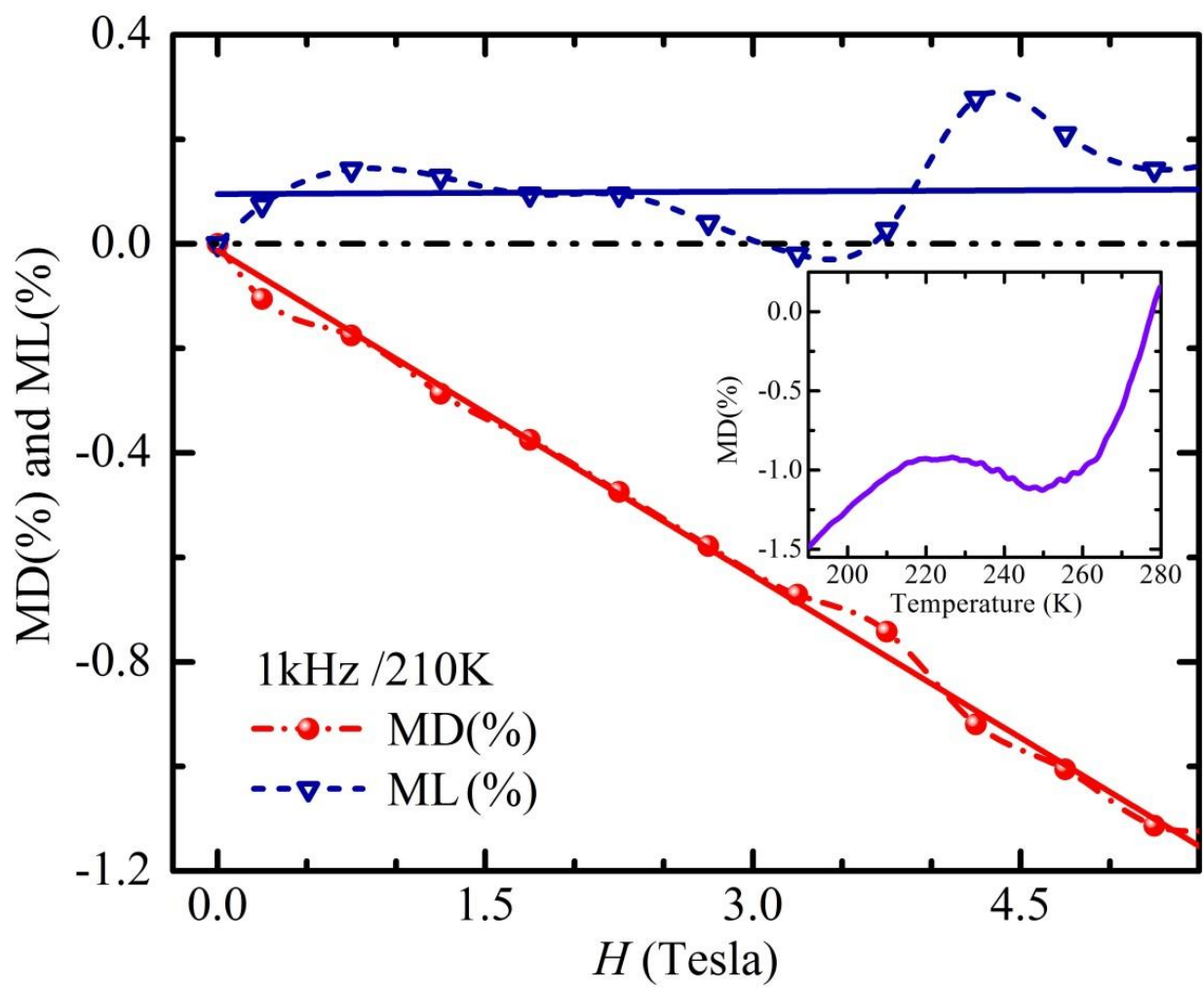

**Fig. 6**